# Discovery of *d*-orbital order in $Tb_2CoAl_4Ge_2$

Zhanyang Hao[1,6*], Haohao Sheng[2,12*], Wanru Ma[3*#], Wengen Zheng[4,13,15*], Zijuan Xie[5*], Wanlin Cheng[3], Zuowei Liang[3], Yongqing Cai[1], Wu Xie[4,13], Junhao Lin[6], Liusuo Wu[6], Zhengtai Liu[7], Mao Ye[7], Ji Dai[8], Massimo Tallarida[8], Shengtao Cui[9], Yogendra Kumar[10], Kenya Shimada[10], Kenichi Ozawa[11], Shuki Torii[14], Kazuhiro Mori[14], Yue Xie[2,12], Junze Deng[2,12], Jiawei Mei[6], Zhenyu Wang[3#], Xianhui Chen[3], Ping Miao[4,13#], Zhijun Wang[2,12#] and Chaoyu Chen[1,6#]

[1] Songshan Lake Materials Laboratory, Dongguan 523808, China.

[2] Beijing National Laboratory for Condensed Matter Physics, and Institute of Physics, Chinese Academy of Sciences, Beijing 100190, China.

[3] Department of Physics, University of Science and Technology of China, Hefei, Anhui 230026, China

[4] Spallation Neutron Source Science Center, Dongguan, 523803, China

[5] International School of Microelectronics, Dongguan University of Technology, Dongguan, 523808, China

[6] Department of Physics, Southern University of Science and Technology, Shenzhen, 518055, China

[7] Shanghai Synchrotron Radiation Facility, Shanghai Advanced Research Institute, Chinese Academy of Sciences, Shanghai, China

[8] ALBA Synchrotron, Carrer de la Llum 2-26 08290 Cerdanyola del Vallès, Barcelona, Spain

[9] National Synchrotron Radiation Laboratory, University of Science and Technology of China, Hefei, China

[10] Research Institute for Synchrotron Radiation Science, Hiroshima University, Higashi-Hiroshima 739-0046, Japan

[11] Institute of Materials Structure Science, High Energy Accelerator Research Organization, Tsukuba, Ibaraki 305-0801, Japan

[12] University of Chinese Academy of Sciences, Beijing 100049, China

[13] Institute of High Energy Physics, Chinese Academy of Sciences, Beijing, 100049, China.

[14] Institute of Materials Structure Science, High Energy Accelerator Research Organization, Tokai, Ibaraki 319-1106, Japan

[15] College of Physics, Nanjing University of Aeronautics and Astronautics, Nanjing, 211106, China

[*] These authors contributed equally to this work.

[#]Correspondence should be addressed to W.M. (mwr19@mail.ustc.edu.cn), Z.Y.W. (zywang2@ustc.edu.cn) P.M. (miaoping@ihep.ac.cn), Z.W. (wzj@iphy.ac.cn) and C.C. (chenchaoyu@sslab.org.cn).

**Abstract:**

**Orbital order describes a quantum state where occupied orbitals line up in a periodic pattern. While orbital physics play a fundamental and universal role in strongly correlated electron systems [1,2], the existence and particularly the band structure fingerprint of orbital order remain a long-standing mystery . Here, we report the discovery of rare earth 5*d*-orbital order developed by the surface states of intermetallic alumogermanide $Tb_2CoAl_4Ge_2$. Angle resolved photoemission spectroscopy (ARPES) reveals characteristic nematic features like Fermi surface deformation and band split, which are decently reproduced by a ferro-orbital order term in the mean-field Hamiltonian. The structural and magnetic origin of such order is excluded by systematic high-resolution neutron powder diffraction (NPD) and scanning tunnelling microscopy (STM) measurements. Uniform orbital polarization from linear dichroism ARPES and spontaneous symmetry breaking domains observed by spatial resolved ARPES and STM conclusively validate the orbital order scenario. Our results not only highlight a methodology that distinguishes orbital physics, but also benchmark a pure *d*-orbital order avoiding complications from structural distortion as in colossal magnetoresistance manganites [1], magnetic order as in iron-based superconductors [3,4], and charge transfer *p*-orbital order in cuprates [5].**

## I. Introduction

Complex solid-state systems owe their razzmatazz to the synergy of lattice, charge, spin and orbital degrees of freedom, bridged through Coulomb repulsion, electron-phonon coupling, exchange, spin-orbital coupling etc. Such synergy results into a wide range of macroscopic phenomena such as magnetism, superconductivity, nematicity, density waves, etc. Described as Landau's spontaneous symmetry-breaking phases, such quantum states of matter are accompanied or even characterized by orders of degree of freedom. For example, charge density wave, charge order, stripe phase and phase separation are dominated by the electronic charge degree of freedom. Similarly, spin density wave, magnetic order, spin fluctuation and skyrmions are rooted in the electronic spin. As the charge and spin manifest as macroscopically measurable quantities, their corresponding orders have been extensively studied, forming the frontiers of modern condensed matter physics.

The third electronic degree of freedom, orbital, and its related quantum order, the orbital order, by contrast, remain largely explored, partially due to the lack of direct experimental probe. By definition, orbital order is a spontaneous symmetry-breaking state in which localized occupied orbitals line up in a periodic pattern, in a similar way as spins do in magnetically ordered structures. On one hand, orbital order is argued to play a fundamental role in shaping the phase diagram and colossal magnetoresistance [6-12] of manganites , the nematicity [13-28] and pseudogap phase [5,29-37] of high-$T_c$ superconductors. On the other hand, the dominant existence of orbital order has never been ascertained in strongly correlated electrons. For example, in perovskite manganites, electronic orbital ordering and structural Jahn-Teller distortion are concurrent [1], rendering it difficult to distinguish the driving force between Kugel-Khomskii superexchange orbital order [38,39] and Kanamori electron-phonon coupling [40]. In Iron-based superconductors, orbital order is theoretically proposed to account for the nematicity, the structural phase transition and the resistivity anomaly [41-47], unfortunately its existence is challenged by the momentum dependence and sign reversal of band splitting [48,49], as directly revealed by ARPES. Additionally, inelastic neutron scattering results [21,27] strongly suggest the spin fluctuation-driven nematicity scenario [3,4]. In cuprates, nematicity emerges at the onset of pseudogap phase [29-37], with signature of orbital order from O $2p$, rather than Cu $3d$, due to charge transfer superexchange [5]. By and large, orbital physics has its ubiquitousness in $3d$ strongly correlated electron systems [1,2,50,51]. Paradoxically, orbital order and its band spectral signature lack definitive recognition, let alone its existence in $4d$ [52,53] and $5d$ systems with extended wave functions.

In this work, we substantiate the existence of $5d$-orbital order by illustrating its clear band structure character, Fermi surface deformation and orbital polarization in rare earth intermetallic alumogermanide $Tb_2CoAl_4Ge_2$. As shown in the phase diagram in Figure 1, bulk $Tb_2CoAl_4Ge_2$ crystals undergo paramagnetic-antiferromagnetic (PM-AFM, $T_{N1}$) and several AFM transitions ($T_{N2}$, $T_{N3}$, $T_{N4}$) at $10\,K < T < 22\,K$. The high-temperature tetragonal to low-temperature orthorhombic structural transition ($T_S$) accompanies the spin reorientation transition ($T_{N3}$) and nematicity-like bulk band split ($T_{Nem}$). The AFM transition $T_{N2}$ occurs with spin density wave (SDW)-like bulk band folding and hybridization gap ($T_{SDW}$). These bulk band features are all captured by density functional theory (DFT) calculations in a single-particle picture. Our key discovery comes from the ARPES observed surface states which develop ferro-orbital order and resultant spontaneous tetragonal symmetry breaking at temperature well above the bulk AFM and structural transition, i.e., $T_{OO} \sim 51\,K \gg T_{N1} = 21.2\,K > T_S = 13.8\,K$. Scanning tunnelling microscopy/spectroscopy (STM/STS) and DFT analyses attribute these surface states to the dangling bonds of surface Tb atoms, with $5d_{xz}/d_{yz}$ orbital origin. Temperature dependent ARPES measurements uncover a surface state nematic transition at $T_{OO}$, characterized by Fermi surface deformation and saddle point in the band structure. STM/STS mapping rules out any structural distortion for the studied surface lattice. Nevertheless, at the surface state bias range, quasi-particle interference (QPI) presents tetragonal symmetry breaking patterns, which can be well matched by scattering vectors between the nematic surface state bands. We model the surface state by constructing a mean-field Hamiltonian containing Pomeranchuk instability and ferro-orbital order interactions. The ferro-orbital order term leads to orbital-dependent band split, decently reproduces the ARPES features. Such orbital order scenario is further validated by the uniform surface state orbital polarization observed in linear dichroism (LD)-ARPES. We also notice nematic domains whose preferential orientations are perpendicular to each other, suggesting the spontaneous symmetry breaking nature. Altogether, our observations establish the emergence of orbital order from rare earth $5d$ electrons, with its existence avoiding the entanglement with the structural distortion as in the colossal magnetoresistance manganites [1], and the AFM order as in the iron-based superconductors [3,4]. Its $d$-orbital origin also distinguishes itself from the $p$-orbital order in cuprates [5], symbolizing pure, strongly correlated orbital physics [2].

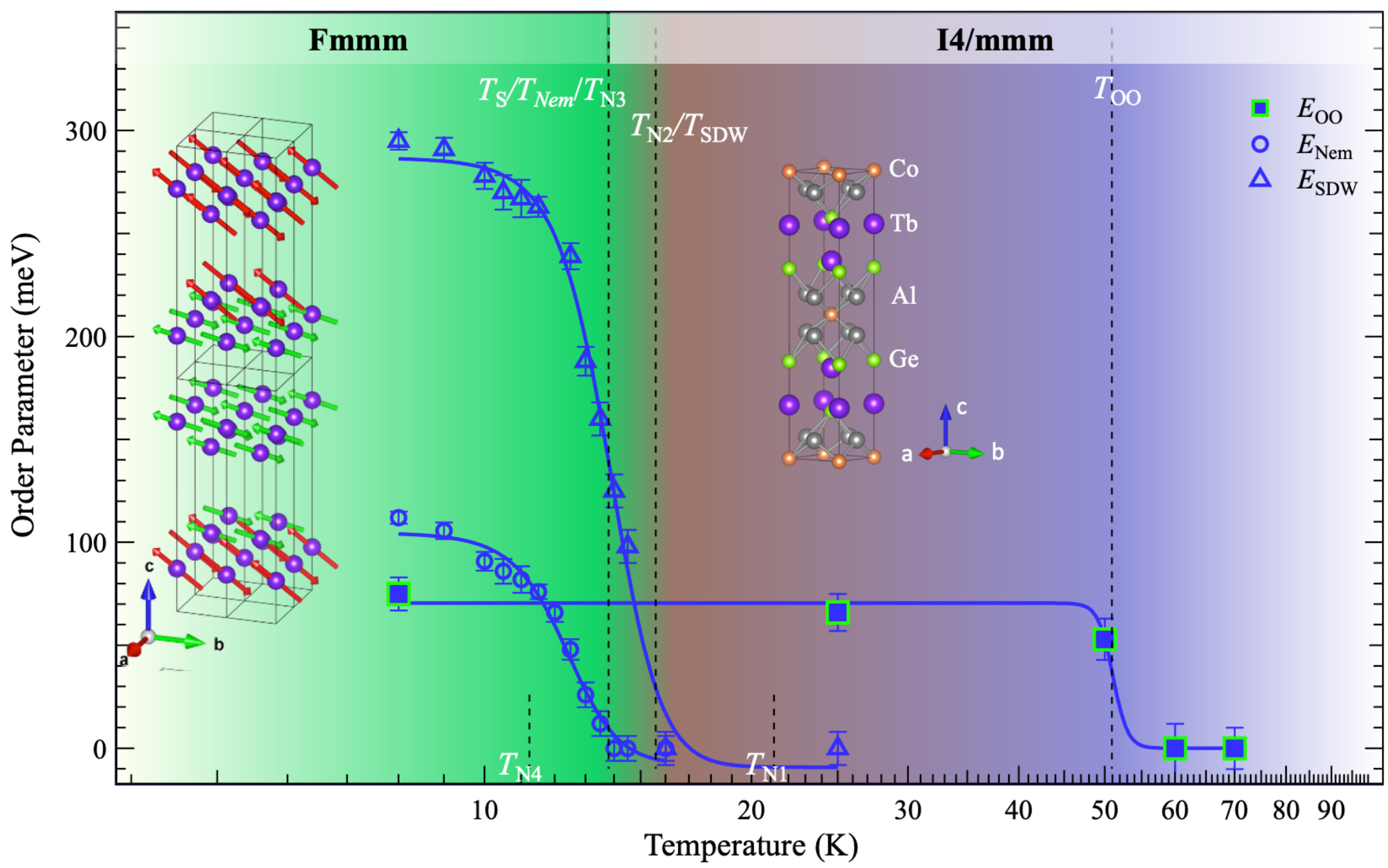

**Figure 1 | Phase diagram of $Tb_2CoAl_4Ge_2$.** NPD, heat capacity and susceptibility measurements reveal four consecutive AFM transitions, at $T_{N1} = 21.2\ K$, $T_{N2} = 15.6\ K$, $T_{N3} = 13.8\ K$ and $T_{N4} = 11.24\ K$, respectively. The insets illustrate the unit cell of magnetic lattice (left) and crystal lattice (right), determined by NPD and X-ray diffraction experiments. Blue symbols and lines are the order parameters, represented by the band splitting extracted from the measured ARPES spectra. Open triangles describe the SDW-like hybridization gap of one bulk band, defining an onset temperature $T_{SDW} = T_{N2} = 15.6\ K$. Open circles represent the nematicity-like energy splitting between the Tb $5d_{xz}$ and $5d_{yz}$ bulk bands, manifesting an onset temperature $T_{Nem} = T_S = T_{N3} = 13.8\ K$, here $T_S$ marks the structural phase transition between high-temperature tetragonal (I4/mmm) and low-temperature orthorhombic (Fmmm) phases, determined by high-resolution NPD experiments. Filled squares indicate the ferro-orbital order induced surface state band anisotropy, with a phase transition temperature $T_{OO} \sim 51\ K$.

## II. Magnetic, structural, symmetry and bulk band evolution

At room temperature, rare earth intermetallic alumogermanide $Tb_2CoAl_4Ge_2$ crystallizes in a tetragonal structure with space group I4/mmm (No. 139, Figure S1, Table S1) [54,55]. As temperature decreases, $Tb_2CoAl_4Ge_2$ experiences 4 consecutive magnetic transitions, which are determined mainly from the intensity change of magnetic Bragg peaks in NPD experiments (Figure S2-4) and corroborated by the slope change of heat capacity and susceptibility (Figure S5 and S6). Curie-Weiss fit of susceptibility (Figure S6) for $300\,K > T > 50\,K$ yields effective moment $\sim 10\,\mu_B$, close to the theoretical value of $Tb^{3+}$. This suggests no moment contribution from Co atoms, which is also confirmed by the magnetic structure analysis on the NPD data (Figures S4 and S7, Tables S3-S6) and related works [56]. The PM-AFM transition occurs at $T_{N1} = 21.2\,K$, indicated by the appearance of a new magnetic Bragg peak in NPD patterns (Figure S2) and slope change in heat capacity (Figure S5), but no clear feature from susceptibility and band structure, probably due to the small magnitude of ordered moments (Figure S4, Table S3). As shown in Figure S1 and S4, the magnetic unit cell of AFM1 is $2 \times 2$ of the tetragonal crystal unit cell and the moments have components along $a$-, $b$- and $c$-axes, with $m_a = m_b \gg m_c$ (Table S3). The magnetic unit cell further doubles itself along the $c$-axis (now $2 \times 2 \times 2$) when entering AFM2 at $T_{N2} = 15.6\,K$, with enlarged moments for the $c$-component (Figure S4, Table S4). Such enlargement validates the magnetic effect on the ARPES observed band structure, manifested as SDW-like bulk band folding and hybridization gap $E_{SDW}$ (Figure 1, $T_{SDW} = T_{N2}$). The major AFM transition dominating the heat capacity and susceptibility curves (Figure S5) occurs at $T_{N3} = 13.8\,K$, accompanied by spin reorientation (AFM3, Figure S4) and concurrent tetragonal-to-orthorhombic crystal structure transition ($T_S$, Figure S1) detected by high-resolution NPD. The orthorhombic crystal structure below $T_S$ is well described by one of the maximal subgroups of I4/mmm, space group Fmmm (No. 69, Figure S1, Table S2). Similar to the nematic phase transition in iron-based superconductors [3,4], the bulk $d_{xz}/d_{yz}$ bands split in energy ($E_{Nem}$) at this temperature, defining a nematicity-like transition with $T_{Nem} = T_S = T_{N3}$. Note that the crystal structure transition not only reduces the in-plane $C_4$ symmetry to $C_2$ but also expands the unit cell approximately to $\sqrt{2} \times \sqrt{2}$ with rotation $45°$ relative to the tetragonal one (Figure S1 and Table S2). On further cooling, the moments reorient again at $T_{N4} = 11.24\,K$ (Figure S4, Table S6), with simultaneous saturation behavior of the SDW-like hybridization gap $E_{SDW}$ and nematicity-like band splitting $E_{Nem}$ (Figure 1). The nature of these two energy order parameters shall be discussed in the following.

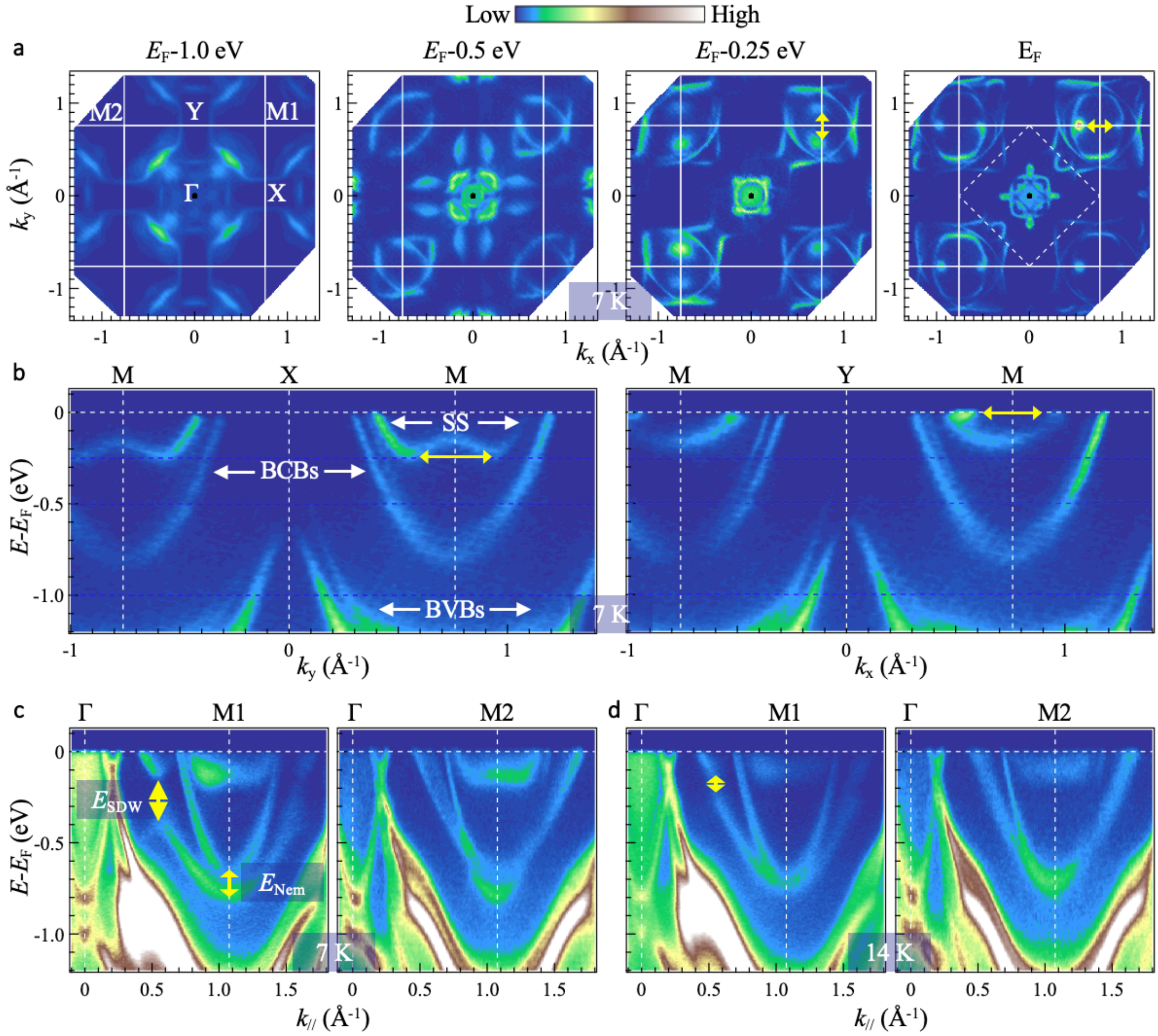

**Figure 2 | Orientation dependent $C_4$ symmetry breaking of surface and bulk electronic states. a**, ARPES spectral intensity in $k_x$-$k_y$ space at various constant energies. White solid lines and characters indicate the high-temperature tetragonal BZ boundaries and high-symmetry points. White dashed lines represent the low-temperature orthorhombic BZ. $E_F$: Fermi level. **b-d**, ARPES band spectra along high-symmetry paths. White arrows and acronyms in **b** show the assignment of surface state (SS), bulk conduction bands (BCBs) and bulk valence band (BVBs). Yellow horizontal arrows in **b** highlight the bottom of the W-shaped surface bands. Yellow vertical arrows in **c** illustrate the order parameters $E_{SDW}$ and $E_{Nem}$, defined as the corresponding band energy split. All the ARPES results here are measured with photon energy 119 eV, at sample temperatures 7 K (**a-c**) and 14 K (**d**), respectively.

Photon energy dependent spectral mappings (Figure S9) exhibit no clear dispersion along $k_z$, suggesting overall two-dimensional electronic structure, in line with the layered lattice structure. We thus only discuss the electronic structure in the cleaved *ab* plane. As shown in Figure 2, the low-energy electronic states of $Tb_2CoAl_4Ge_2$ consist of steeply dispersed band manifolds around the Brillouin zone (BZ) center Γ (Figures 2a and 2c), and several electron-type bands enclosing the BZ corner M points (Figures 2a and 2b). In the present work, we concentrate on these M-region bands, including two parabolic ones with deep bottoms at $\sim -0.8\ eV$, and one W-shaped band with shallow bottoms at $\sim -0.25\ eV$ (Figure 2b, left). According to our DFT calculations (Figures S10-S11), the formers are bulk conduction bands (BCBs) while the latter is dangling bond surface state residing on the double-Tb-layer termination, both from Tb $5d_{xz}/d_{yz}$ orbitals.

The constant energy contours at various binding energies and high-symmetry-path spectra measured at 7 K (Figures 2a-2c) exhibit tetragonal symmetry breaking for both bulk and surface bands. At high binding energy 1.0 eV (Figure 2a, left), the bulk valence bands (BVBs) obey the in-plane $C_4$ symmetry. Moving to $E_F$ (Figure 2a, right), the BCBs break the $C_4$ symmetry by developing band folding features with respect to the orthorhombic BZ border (white dashed lines). The main and folded outer BCBs hybridize at this border, which is the middle point of $\Gamma - \mathrm{M1}$ of the tetragonal BZ, and open an SDW-type gap ($E_{SDW}$, Figure 2c, left). On the contrary, along $\Gamma - \mathrm{M2}$ the folding and gap cannot be resolved in ARPES spectra (Figure 2c, right), which is attributed to the minute gap size according to DFT calculation (Figure S12). We also observe band splitting between the $5d_{xz}/d_{yz}$ BCBs along $\Gamma - \mathrm{M1/M2}$, reminiscent of the $3d_{xz}/d_{yz}$ nematic splitting in Fe-based superconductors [49], thus defining an order parameter $E_{Nem}$.

The evolution of order parameters $E_{Nem}$ and $E_{SDW}$ with temperature can be tracked by measuring their magnitudes in temperature dependent ARPES spectra. As shown in Figure 2d, at temperature $T_{N3} < T = 14\ K < T_{N2}$, $E_{SDW}$ reduces significantly but remains visible, yet $E_{Nem}$ vanishes. Systematic temperature dependent analyses in Figures S13-S14 show that, $E_{SDW}$ appears below 16 K, with estimated $T_{SDW} = T_{N2} = 15.6\ K$. $E_{Nem}$ emerges with the onset of AFM3 and structural transition, *i.e.*, $T_{Nem} = T_S = T_{N3}$, again, imitating the magneto-elastic coupling and nematicity in some iron-based superconductors [3,4]. Nevertheless, the behaviors of $E_{SDW}$ and $E_{Nem}$ can be reproduced in our DFT calculation (Figure S12) by including the corresponding magnetic structure, suggesting AFM origin of the bulk band $C_4$ symmetry breaking.

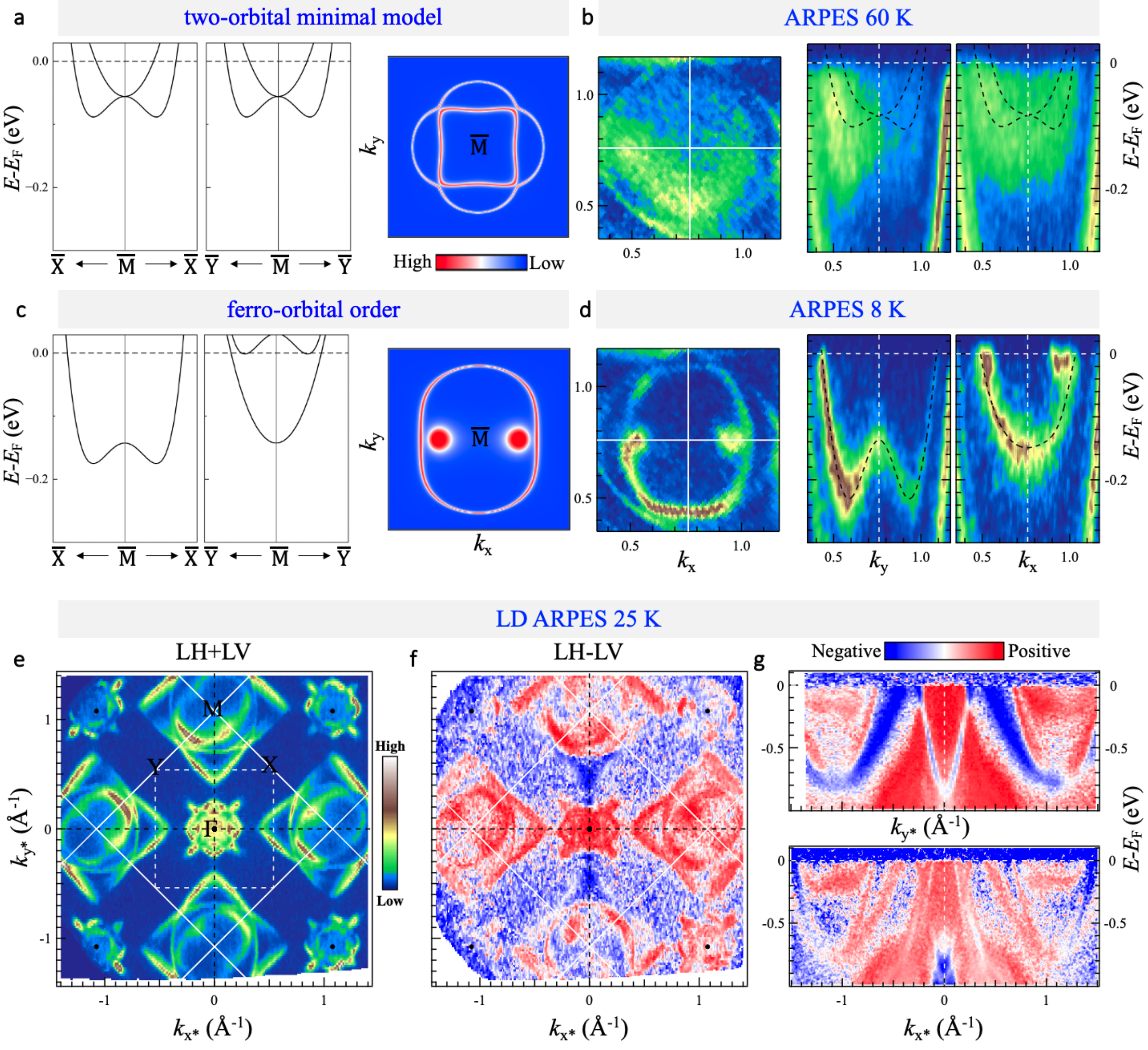

**Figure 3 | Surface ferro-orbital order characterized by Fermi surface deformation, nematic orbital split and orbital polarization. a, c**, Results of two-orbital minimal model. Surface-state energy bands and Fermi surface near $\overline{\mathrm{M}}$ point calculated from the model (**a**) without and (**c**) with ferro-orbital order. $v = 0.36\ eV$ and $g = 0$. **b, d**, Corresponding ARPES results. Fermi surface and surface state spectra overlaid with extracted dispersions (black dashed lines) measured at sample temperature 60 K (**b**) and 8 K (**d**). (**e**) ARPES Fermi surface integrating photoemission intensity from photons with LH and LV polarization. (**f**) LD Fermi surface. (**g**) Corresponding LD spectra along $k_{y*}$ (top) and $k_{x*}$ (bottom) directions. In **d, e, f,** we use the low-temperature orthorhombic BZ notations. Sample temperature is 25 K.

## III. Surface ferro-orbital order and orbital polarization

The surface states manifest distinct symmetry breaking at low temperature with elongated Fermi surfaces along $\mathrm{X}-\mathrm{M1}$ (Figure 2a and 3d). We construct a Wannier tight-binding model from the nonmagnetic calculations, then use an iterative method to obtain the surface Green's function of the semi-infinite system. The experiment surface is found to be terminated on the 2-Tb layers (see Figure S11). The corresponding Fermi surface projections present surface state contours resembling the "Chinese-knot" (Figure S11c). In general, the bulk and surface Fermi surface projections (Figure S11c and S11f) are consistent with the ARPES results (Figure 3b, 60 K)

To simulate the evolution of surface states, we build a two-orbital minimal model, which reads,

$$H_0(k_x,k_y) = m\sigma_0 + A(k_x,k_y)\sigma_0 + B(k_x,k_y)\sigma_z + C(k_x,k_y)\sigma_x, \qquad (1)$$

with $A(k_x,k_y) = t_1(k_x^2+k_y^2) + t_1'(k_x^4+k_y^4) + t_1''k_x^2k_y^2$, $B(k_x,k_y) = t_2(k_x^2-k_y^2) + t_2'(k_x^4-k_y^4)$ and $C(k_x,k_y) = t_3k_xk_y + t_3'(k_x^3k_y + k_xk_y^3)$. Such 2D model can reproduce the DFT surface band structure and $C_{4V}$ symmetry well (see the parameters and results in Figure S15 and Table S7). As shown in Figure 3a, the two surface state bands degenerate at $\bar{M}$ point and the Fermi surfaces present "Chinese-knot" contours with $C_4$ symmetry, capturing the ARPES observed features at 60 K (Figure 3b).

Regarding surface state symmetry breaking, we consider the following interaction Hamiltonian,

$$H_{int} = \frac{S}{2(2\pi)^2}\iint d^2kd^2k'[f_{kk'}n_k^+n_{k'}^+ - vn_k^-n_{k'}^-]. \qquad (2)$$

Here, $f_{kk'} = -gd_kd_{k'}$ and $d_k = k_x^2 - k_y^2$. $n_k^{\pm} = n_{xk} \pm n_{yk}$. $S$ is the area of the unit cell. $g$ and $v$ are the interaction strengths. $n_{xk}$ and $n_{yk}$ are density operators of $d_{xz}$ and $d_{yz}$ orbitals, respectively. In Eq. (2), the first term corresponds to Pomeranchuk instability via forward-scattering interaction [57,58], while the second term represents ferro-orbital order from Kugel-Khomskii interaction [59,60] at mean-field level without magnetization. Then, we solve the total Hamiltonian $H_0 + H_{int}$ by the Hartree-Fock meanfield method. Without loss of generality, we consider the two orderings individually. The results of $d$-wave Pomeranchuk instability are shown in Figure S15, whose Fermi surfaces are still Chinese-knot like with weak $C_4$ breaking.

The results of the ferro-orbital order (Figure 3c) can explain the low-temperature ARPES data (Figure 3d) very well. First, the double degeneracy of surface bands at $\bar{M}$ is lifted. Second, the $C_4$ symmetry is broken clearly from the simulated dispersions along $\bar{M}-\bar{X}$ and $\bar{M}-\bar{Y}$. Third, the constant energy contours decently reproduce the ARPES Fermi surface. Particularly, the two point-shaped features come from the bottoms of the W-shaped band located slightly above the Fermi level. Based on these agreements, we conclude that the surface state nematic symmetry breaking originates from the ferro-orbital order in $Tb_2CoAl_4Ge_2$.

We trace the ferro-orbital order transition through the temperature dependence of surface states. As shown in Figures 3b, 3d and S16, Fermi surfaces remain elongated up to 50 K and regain the $C_4$ symmetry only reaching 60 K. At 8 K (Figures 3d and S16b top), the surface spectra along $k_y$ is W-shaped, with global band peak minima at $\sim -0.22\ eV$ and local maxima (M point) at $\sim -0.14\ eV$. Changing to $k_x$, it is U-shaped with minima at $\sim -0.14\ eV$, suggesting a saddle point formed at M. At 60 K, the spectral anisotropy disappears (Figures 3b and S16b bottom). Defining the energy difference between the W-U band minima as the ferro-orbital order parameter $E_{OO}$, we track its temperature evolution (Figure S16d-e). As summarized in Figure 1, $E_{OO}$ sets on roughly at $\sim 51\ K$, more than twice higher than the bulk AFM and structure transition, $T_{OO} \sim 51\ K \gg T_{N1} = 21.2\ K > T_S = 13.8\ K$, suggesting the ferro-orbital order is decoupled from the bulk AFM order and orthorhombic structure transition.

Analogous to spin polarization as the characteristic of ferromagnetic order, orbital polarization directly evidences ferro-orbital order [11,13,61]. We employ orbital sensitive LD-ARPES, in which the selection rules demand simultaneous odd or even symmetry to the mirror plane for initial state orbitals and beam polarization vector. Here we switch to the orthorhombic BZ notations, in which $x^*z^*$ and $y^*z^*$ define the mirror planes. Shown in Figures 3e-f are the sum and difference, respectively, of the Fermi surfaces mapped using linear horizontal (LH) and linear vertical (LV) polarization. In the LH+LV Fermi surface (Figure 3e), spectral intensity of the main features obeys the $C_4$ rotation symmetry. While in the LD (LH-LV, Figure 3f) Fermi surface, the outer BCBs surrounding M points show negative dichroism along $k_{y*}$, suggesting $d_{xz}$ orbital which is odd to the $y^*z^*$ measurement plane. Conversely, along $k_{x*}$, the dichroism is positive, suggesting $d_{yz}$ orbital. Such *d*-wave orbital polarization in reciprocal space is reminiscent of the spin polarization in our recently discovered *d*-wave altermagnet [62]. Whether a *d*-wave alter-orbital order exists, and its relevant physics remain to be explored.

As expected, the surface state bands and Fermi surfaces show uniform positive dichroism spanning the whole reciprocal space, suggesting dominant $5d_{yz}$ orbital occupation and full orbital polarization. As the sample is in the paramagnetic phase (25 K) for LD-ARPES, such orbital polarization is uniquely associated to and serving as a direct physical consequence of the surface ferro-orbital order.

## IV. Surface intra-unit-cell orbital order

To further validate the origin of the surface states and the ferro-orbital order induced nematicity, we employ spectroscopic-imaging STM, which detects electronic wave function in real and reciprocal space simultaneously. Structurally, $Tb_2CoAl_4Ge_2$ crystals could cleave between either adjacent Ge and Tb layers or two Tb layers, exposing two types of Tb terminations (Figure 4b, Tb1 and Tb2). Figure 4a shows the typical atom-resolved topography achieved from four cleaved samples; a comparison of the relative heights of multiple consecutive steps in a larger field of view (FOV, Figure 4b) to the schematic crystal structure suggests that the large and flat surfaces are the Tb1 terminations. This is further supported by our DFT calculations (Figure S11). Interestingly, the topographic images obtained at positive biases exhibit a marked $C_4$-symmetric square lattice of Tb atoms (note that the difference between lattice constants *a* and *b* in the orthorhombic phase is too weak to be distinguished), while at negative biases a striped pattern appears, breaking the $C_4$ symmetry within one unit cell (Figure S17). Since a topographic image contains convoluted information on both lattice corrugation and electronic density modulation, and the former is, in principle, independent of the tunnel bias, the observed striped pattern is most likely to be a manifestation of rotation symmetry breaking of the underlying electronic state.

To get a clearer picture of any such electronic order, we image the energy-resolved local density of states $g(\boldsymbol{r}, eV)$ in the same FOV of Figure 4a. Figures 4c-d display two such examples at 200 and -100 meV, respectively, with their Fourier transforms (FT) shown in the insets. Again, the electronic modulation seen in $g(\boldsymbol{r}, 200\ \text{meV})$ is rather $C_4$-symmetric with roughly equal intensities at Bragg wavevectors $Q_x = (\pm 1, 0)\ 2\pi/a$ and $Q_y = (0, \pm 1)\ 2\pi/a$ (Figure 4e). Contrarily, $g(\boldsymbol{r}, -100\ \text{meV})$ clearly demonstrates a unidirectional modulation along one of the two lattice directions, and its FT (Figure 4d inset) exhibits inequivalent intensities at $Q_x$ and $Q_y$ (Figure 4f). Tracking the intensity of the Bragg peaks as a function of energy, we find the peak amplitudes exhibit a markedly different energy dependence (Figure 4g). We define a normalized order parameter, $\mathrm{Z(eV)} = (\mathrm{M}(\mathrm{Q_Y}) - \mathrm{M}(\mathrm{Q_X}))/(\mathrm{M}(\mathrm{Q_Y}) + \mathrm{M}(\mathrm{Q_X}))$ to measure the anisotropy of electronic modulations between $Q_x$ and $Q_y$. The plot of $\mathrm{Z}(eV)$ in Figure 4h reveals prominent anisotropy in the bias window of -0.25 to 0.05 V, matching the surface state energy range. Furthermore, a domain boundary at which the striped pattern rotates by 90° has been revealed on the same surface (Figure S17), excluding the STM tip anisotropy and local strain effect as possible origin of this $C_2$ modulation. These findings are a strong indicator that the electronic states on the Tb1 terminations spontaneously break the $C_4$ symmetry at low temperatures.

Correspondingly, the electronic structure in reciprocal space should also exhibit $C_4$ symmetry breaking at the same energy scale. To test this assumption, we study the electronic band structure through QPI imaging. Figure 4i shows the FT of a selected image at $g(\boldsymbol{r}$, -10 meV) obtained in a larger FOV, in which a rich array of scattering events can be identified with a comparison to the ARPES Fermi surface (Figure 4j). While the intra-band scattering of bulk $\alpha$ pocket preserves $C_4$ symmetry, the surface related scattering patterns develop clear nematicity. As shown by the colored arrows, the arc features ($q_2$) come from scattering between the elliptical surface state Fermi surfaces, while point features ($q_3$) connects the bright points formed by the bottoms of one surface band, which is pushed up over the Fermi level by the ferro-orbital order (see Figure 3). Both $q_2$ and $q_3$ are anisotropy between $x$ and y direction, while $q_1$, scattering between intra-band bulk Fermi surfaces, is $C_4$ symmetric. These results not only establish an intra-unit-cell electronic nematicity free from any bulk lattice distortion but also attribute such spontaneous nematic symmetry breaking to the surface ferro-orbital order.

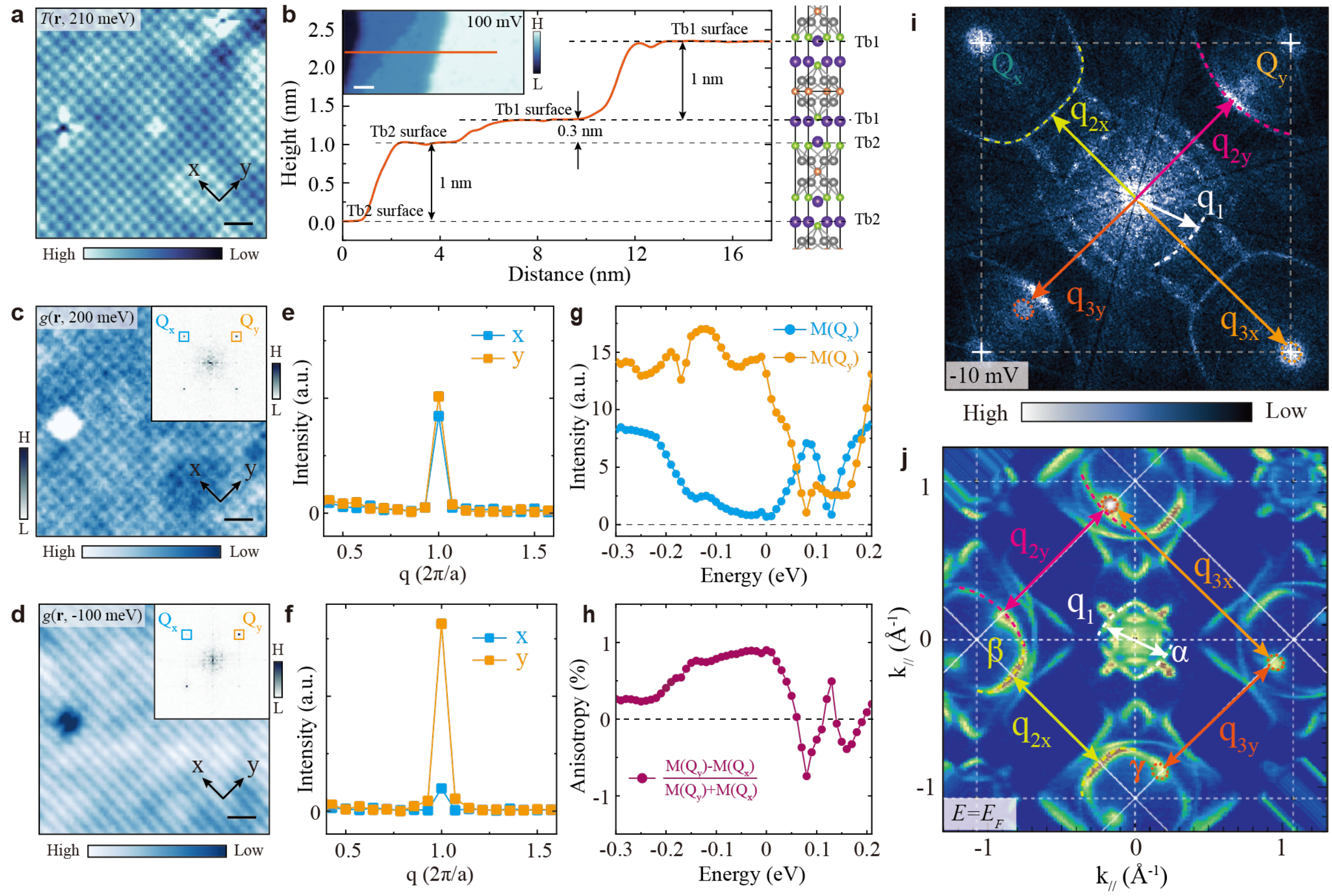

**Figure 4 | Surface electronic nematicity originated from the Tb termination of $Tb_2CoAl_4Ge_2$. a**, Atomically resolved topographic image. **b**, Height profile obtained along the orange line in the inset (STM topography) showing that the main cleavage planes are Tb1 terminations. **c**, **d**, Typical d$I$/d$V$ map acquired at 200 meV (**c**) and -100 meV (**d**) from the FOV in **a**. The inset shows the corresponding FT. **e, f**, Intensities of the Bragg points from **c** and **d**, respectively. **g**, The intensity of atomic lattice in the FT, M($Q_x$) and M($Q_y$), as a function of energy (STM bias). **h**, Anisotropy ratio of the intensities $Z(eV)$. **i**, The QPI pattern showing multiple scattering features, labelling wave vectors as colored arrows along $Q_x$ and $Q_y$. **j**, Fermi surface map. The colored double arrows indicate the corresponding scattering patterns in **i**. Setup conditions: sample bias $V_S$ = 210 mV, tunneling current $I_t$ = 4 nA (**a**), $V_S$ = 100 mV, $I_t$ = 600 pA (**b** inset), $V_S$ = 200 mV, $I_t$ = 4 nA, bias modulation $V_m$ = 4 mV (**c**), $V_S$ = -100 mV, $I_t$ = 4 nA, $V_m$ = 4 mV (**d**). Scale bars: 1 nm (**a**, **c**, **d**), 3 nm (**b** inset).

## V. Discussion

In strongly correlated 3*d* electron systems, orbital order has been long proposed to account for or be intimately related to a cascade of phenomena such as Verwey transition [50,63,64], cooperative Jahn-Teller distortion [65,66], colossal magnetoresistance [9,10,12,67], nematicity [41-47] and the pseudogap phase[5]. Nevertheless, the decisive fingerprint of orbital order is unclear. The reasons are two-fold. In essence, orbital is frequently entangled with lattice, charge and spin degrees of freedom, so that the development of one order immediately induces the others, making the origin of such spontaneous symmetry breaking a physics realization of the 'chicken and egg problem' [1,3,4]. Methodologically, unlike spin and charge orders which are accessible directly via magnetization or polarization, there lacks an acknowledged approach to identify the characteristic of orbital order. Intuitively, the ordering temperature of the orbital we observed here is more than twice higher than that of AFM and structural distortion, distinguishing dominant orbital physics. Its 5*d* orbital origin also expanding the arena where correlated orbital physics could emerge. On a deeper level, not only we have pointed out the electronic features of orbital order as Fermi surface deformation and band orbital splitting in ARPES, but also the unique orbital polarization is revealed via LD-ARPES. Furthermore, spatial resolved ARPES uncovers nematic domains with preferential orientation parallel for the bulk states but mutually orthogonal for the surface states (Fig. S18), in line with the STS map (Fig. S17), confirming the spontaneous symmetry breaking nature of orbital order. These approaches, combining different facets of ARPES functionality, set a new benchmark against which one can measure the intricate role that orbital physics is playing in a variety of phenomena through its strong coupling with charge, spin, and lattice dynamics.

## VI. Methods

### Sample growth and characterization

$Tb_2CoAl_4Ge_2$ single crystals were grown by the Al flux method. High-purity Tb (chunks), Co (powder), Al (rods), and Ge (chunks) were mixed in a ratio of 2:1:18:2 and placed into an alumina crucible. The crucible was then placed into a quartz tube and vacuum sealed. The tube was first heated to 1050°C and held for 12 hours. Then, the temperature was lowered to 750°C in 150 hours and held there for 5 hours. Afterward, the quartz tube was placed into a centrifuge to separate the flux from the single crystals. Millimeter-sized tetragonal single crystals were obtained.

The structure of the crystals was characterized by X-ray diffraction with Cu Kα radiation using a Rigaku Smartlab 9kW diffractometer. Heat capacity and magnetic susceptibility measurements were performed using Physical Property Measurement System (PPMS) and Magnetic Property Measurement System (MPMS3) of Quantum Design, respectively.

Temperature dependent high-resolution NPD experiments were conducted on the high-resolution neutron diffractometers TREND of China Spallation Neutron Source [68] and SuperHRPD of Japan Proton Accelerator Research Complex [69]. The powder crystalline samples, ground from the single crystals, were mounted in top-loading cryofurnaces and measured in the temperature range of 3.7-300 K. The collected neutron patterns were analyzed using the FELLPROF suite [70] and JANA2020 program [71].

### ARPES measurements

ARPES measurements were performed at the BL03U beamline of the Shanghai Synchrotron Radiation Facility, the BL1 beamline of Research Institute for Synchrotron Radiation Science, Hiroshima University, the BL13U of National Synchrotron Radiation Laboratory, Hefei, the LOREA beamline of ALBA Synchrotron and the BL28A beamline of High Energy Accelerator Research Organization, Tsukuba. The samples were cleaved in situ under base pressure better than $5\times10^{-11}$ mbar and temperature below 15 K. The energy resolution was set to 10-20 meV depending on the photon energy used, and the angular resolution was set to $0.2^{\circ}$.

### First-principles calculations

We carried out first-principles calculations based on density functional theory (DFT) with the projector augmented wave (PAW) method [72,73], as implemented in the Vienna *ab initio* simulation package (VASP) [74,75]. The generalized gradient approximation (GGA) in the form of the Perdew-Burke-Ernzerhof (PBE) functional [76] was employed for the exchange-correlation potential. The kinetic energy cutoff for the plane-wave expansion was set to 400 eV. The Brillouin zone was sampled by the Monkhorst-Pack method in the self-consistent process, with a $10\times10\times10$ **k**-mesh for nonmagnetic phase and a 4×4×2 **k**-mesh for antiferromagnetic (AFM) phase. The PBE+U method [77] was considered for AFM calculations, where the parameter Hubbard U is selected to be 7 eV for Tb 4*f* orbitals. The band structure of the AFM phase was obtained using the band unfolding method, as implemented in the VASPKIT package [78]. The maximally localized Wannier functions were constructed using the Wannier90 package [79]. The surface electronic spectra and the Fermi surface were calculated using surface Green's function of the semi-infinite system [80,81], as implemented in the WannierTools package [82].

**Scanning tunnelling microscopy measurements**

The STM/STS experiments were performed with a commercial CreaTec low-temperature STM system. $Tb_2CoAl_4Ge_2$ single crystals were cleaved under ultrahigh-vacuum conditions ($1\times10^{-10}$ Torr) at ~30 K and then immediately inserted into the STM head for measurements. The data were acquired at 4.5 K unless otherwise specified, using PtIr tips that were pretreated and calibrated on Au (111) surfaces prior to measurements. Spectroscopic data were recorded using standard lock-in technique with an AC modulation voltage of 1-4 mV at the frequency of 987.5 Hz. The Lawler–Fujita drift-correction algorithm[32] was employed to remove the drift effects in the spectroscopic maps, and the resulted FTs are two-fold symmetrized to increase the signal-to-noise ratio.

**Data availability**

All data are available in the main text or the supplementary materials. Further data are available from the corresponding author on reasonable request.

**ACKNOWLEDGEMENTS**

This work is supported by the National Key R&D Program of China (Grants No. 2022YFA1403700 and No. 2022YFA1403800), the National Natural Science Foundation of China (NSFC) (Grants No. 12574068 and No.12188101), Guangdong Basic and Applied Basic Research Foundation (Grants No. 2022B1515020046, 2022B1515130005, 2021B1515130007, 2022B1515120014, 2023B0303000003, and 2023B1515120060), the Guangdong Innovative and Entrepreneurial Research Team Program (Grants No. 2019ZT08C044 and 2021ZT09C539), Shenzhen Science and Technology Program (Grant No. KQTD20190929173815000) and the center for materials Gemonme. The authors also acknowledge the neutron beam time at TREND (https://cstr.cn/31113.02.CSNS.TREND) of CSNS (https://cstr.cn/31113.02.CSNS), the neutron beam time (Proposal No. 2023BF0801) at SuperHRPD of J-PARC and the Shanghai Synchrotron Radiation Facility (SSRF) of BL03U(31124.02.SSRF.BL03U) for the assistance on ARPES measurements.

**Author contributions**

C.C. conceived and supervised the research. Z.H., Y.C., Z.L., M.Y., J.D., M.T., S.C., Y.K., K.S. and K.O. contributed to the ARPES instruments, measurement and analysis. Z.H., Z.X., Y.C., J.M., J.L., and L.W. contributed to the sample growth and characterization. W.Z., W.X., S.T., K.M. and P.M. performed the NPD measurements and analysis. H.S., Y.X., J.D. and Z.W. performed calculations/simulations. W.M., W.C. Z.Y.W. and X.C. performed the STM/STS measurements and analysis. All authors wrote and corrected the manuscript.

**Competing interest**

The authors declare no competing interests.

**Correspondence and requests for materials** should be addressed to Chaoyu Chen